\documentclass[aps,reprint,prl,twocolumn,superscriptaddress]{revtex4-2} 
\usepackage{amsfonts}
\usepackage{amssymb}
\usepackage{amsthm,amsmath}
\usepackage{mathrsfs}
\usepackage{indentfirst}
\usepackage{multirow}
\usepackage{graphicx}
\usepackage{amsmath}
\usepackage[amssymb]{SIunits}
\usepackage{amssymb}
\usepackage{color}
\usepackage[usenames,dvipsnames]{xcolor}
\usepackage{natbib}
\usepackage[T1]{fontenc}
\usepackage[hidelinks, bookmarks=true]{hyperref}
\usepackage{bm}
\usepackage{booktabs}
\usepackage{multirow}
\usepackage{setspace}

\begin{document}

\title{Melting Ice Activates Mechanical Forcing from Entrapped Air}

\author{Jin Hu}
\affiliation{New Cornerstone Science Laboratory, Center for Combustion Energy, Key Laboratory for Thermal Science and Power Engineering of MoE, Department of Energy and Power Engineering, Tsinghua University, Beijing 100084, China}

\author{Feng Wang}
\email{fengwang2023@tsinghua.edu.cn}
\affiliation{New Cornerstone Science Laboratory, Center for Combustion Energy, Key Laboratory for Thermal Science and Power Engineering of MoE, Department of Energy and Power Engineering, Tsinghua University, Beijing 100084, China}

\author{Yihong Du}
\affiliation{New Cornerstone Science Laboratory, Center for Combustion Energy, Key Laboratory for Thermal Science and Power Engineering of MoE, Department of Energy and Power Engineering, Tsinghua University, Beijing 100084, China}

\author{Enrico Calzavarini}
\affiliation{Unit\'e de M\'ecanique de Lille--J. Boussinesq (UML) ULR 7512, Universit\'e Lille, F-59000 Lille, France}

\author{Chao Sun}
\email{chaosun@tsinghua.edu.cn}
\affiliation{New Cornerstone Science Laboratory, Center for Combustion Energy, Key Laboratory for Thermal Science and Power Engineering of MoE, Department of Energy and Power Engineering, Tsinghua University, Beijing 100084, China}
\affiliation{Department of Engineering Mechanics, School of Aerospace Engineering, Tsinghua University, Beijing 100084, China}

\date{\today}

\begin{abstract}
Natural ice commonly contains entrapped air, yet its dynamical role during melting remains unresolved. Using numerical simulations and theoretical analysis, we show that melt-front-released bubbles activate entrapped air as a mechanical forcing on the liquid. Bubble buoyancy disrupts density-anomaly-induced thermal structure and injects momentum into the melt-controlling boundary layer, producing melt rates more than twice the bubble-free value at percent-level gas fractions. The evolving interface and background convection regulate this coupling, which weakens again under sufficiently vigorous convection, producing a finite coupling window. A mechanical-power ratio comparing bubble-generated and background-flow power organizes this behavior. Recasting the bubble contribution as an effective near-boundary velocity extends standard ice-ocean melt closures, while reported field conditions can enter the mechanically significant regime. More broadly, a moving phase boundary can activate an embedded dispersed phase as an internal mechanical forcing.
\end{abstract}

\maketitle




Melting at ice-water boundaries is a canonical moving-boundary problem whose dynamics can be strongly modified by the surrounding flow~\cite{du2024physics,YangR2023,anderson2020convective,worster1997convection,Weady2022AnomalousConvective}. This coupling is central to glaciers, ice shelves, and sea ice, with consequences for freshwater input, ocean circulation, and sea level~\cite{roach2025physics,cenedese2023icebergs,hewitt2020subglacial,rosevear2021role,washam2023direct}. In standard ice-ocean parameterizations, mechanical forcing enters through the near-boundary velocity: it determines a friction velocity and turbulent exchange rates, while heat, salt, and freezing-point constraints close the interfacial melt problem~\cite{HOLLAND1999ModelingThermodynamic,MCPHEE2008RevisitingHeat,JENKINS2010ObservationParameterization,MCCONNOCHIE2017TestingCommon}. Field measurements confirm that characterizing the boundary-layer state accurately is essential~\cite{NASH2024TurbulentDynamics,WEISS2025DirectMeasurement}. Yet these formulations do not account for mechanical forcing that is activated by the melting ice itself.

Natural ice provides such a source because it commonly contains air pockets and bubbles~\cite{Gow1975,Thievenaz2025,Meijer2024}, with typical sizes ranging from submillimetric to millimetric scales and gas fractions reaching the percent level~\cite{Gow1975,Wengrove2023}. Yet their influence on melting is not determined by gas content alone. Early experiments with manufactured gas-infused ice reported little measurable effect on melt-driven convection, whereas later measurements on natural brash ice and glacier ice found appreciable melt enhancement in bubbly samples~\cite{josberger1980effect,smith1996study,Wengrove2023}. In particular, glacier-ice experiments showed that bubble release can strongly agitate the boundary layer and approximately double the natural-convection melt rate~\cite{Wengrove2023}. In these glacier-ice experiments, however, pressure-driven ejection, gas expansion, buoyant rise, salinity effects, and melt-driven convection occur simultaneously, leaving the mechanical contribution of post-release bubble motion unresolved.

A key distinction is that bubbles released by a melting interface differ fundamentally from bubbles injected externally into an established flow~\cite{2025LEGENDREGasBubble,mathai2020bubbly,gvozdic2018experimental,gvozdic2019experimental}. Their release locations are selected by the instantaneous phase boundary, their release rate is coupled to melting, and their trajectories are filtered by the evolving interface geometry and surrounding circulation. Melt-front release therefore creates a distinct coupling problem in which the motion of the boundary releases a dispersed phase whose subsequent dynamics can in turn modify interfacial transport. How this coupling converts entrapped air into an effective mechanical forcing on the melt boundary layer remains unresolved.

In this Letter, we isolate this feedback and show that melting activates entrapped air as a state-dependent mechanical forcing. The evolving interface and background convection jointly regulate how the buoyancy of released bubbles is converted into motion that perturbs the melt-controlling boundary layer, producing changes in both thermal structure and interfacial transport. These dynamics distinguish the availability of entrapped gas from its dynamical effectiveness and lead to a mechanical-power criterion that separates bubble-generated forcing from the response of the background flow. The criterion identifies when the internally generated forcing becomes dynamically significant, while recasting the bubble power as an effective near-boundary velocity provides a direct route for incorporating this effect into existing ice-ocean melt models.

\begin{figure*}[!t]
\centering
\includegraphics[width=1\textwidth]{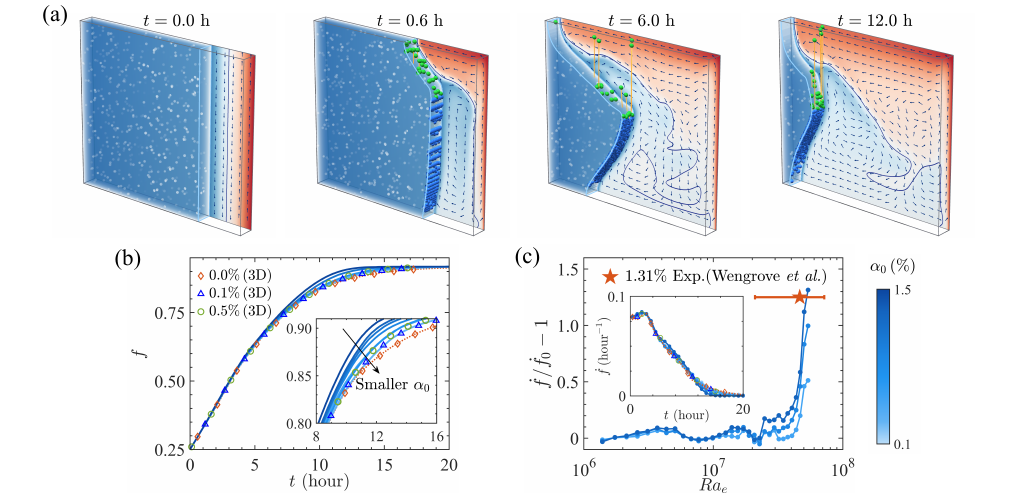}
\caption{\textbf{Release of entrapped air activates an internal forcing during melting.}
\textbf{(a)} Representative 3D evolution for $\alpha_0=0.5\%$ at $t=0$, $0.6$, $6$, and $12\,\mathrm{h}$. White particles are bubbles still trapped in ice. Released bubbles follow the evolving liquid pathways. Larger green and blue filled circles denote released bubbles originating above and below the interface turning point, respectively. Bubble sizes are exaggerated for clarity and are not to scale.
The melt front, the $4\,^{\circ}\mathrm{C}$ isotherm, and the liquid velocity reveal the reorganization of the convective state.
\textbf{(b)} Global liquid fraction $f$ for bubble-free and bubbly cases. Lines are 2D results and symbols are 3D results. The inset enlarges the late stage.
\textbf{(c)} State-matched melt-rate enhancement, $\dot f/\dot f_0-1$, versus $Ra_e=Ra\,f^3$, where $\dot f_0$ is the bubble-free value at the same $f$. The orange star is the glacier-ice experiment of Ref.~\cite{Wengrove2023}, included only as a regime-level magnitude reference. Inset: the corresponding time-based melt rates.}
\label{Fig-1}
\end{figure*}

To isolate this mechanism, we consider convective ice melting in a sidewall-heated square cavity of height $H=0.1\,\mathrm{m}$, a canonical freshwater configuration in which the density maximum near $4\,^{\circ}\mathrm{C}$ strongly reorganizes the melt-driven circulation~\cite{wang2021ice,hu2024numerical}. The cold sidewall is maintained at $T_c=-1\,^{\circ}\mathrm{C}$, while the horizontal walls are adiabatic. The reference hot-wall temperature is $T_h=10\,^{\circ}\mathrm{C}$, which lies within the upper range observed in summer polar meltwater bodies~\cite{Carrea2025,Young2024}. Additional simulations at $T_h=15$ and $20\,^{\circ}\mathrm{C}$ are used to probe progressively stronger background convection. Initially, the ice-water interface is vertical, with global liquid fraction $f(0)=0.25$. Air is entrapped in the ice at volume fractions $\alpha_0=0.1\%$-$1.5\%$ and is released as the melt front advances [Fig.~\ref{Fig-1}(a)]. Released bubbles are modeled as small, spherical, low-Stokes-number point particles, with bubble-to-fluid momentum feedback retained. Details of the numerical model, resolution, and convergence tests are given in End Matter and the Supplemental Material.

Figure~\ref{Fig-1}(a) shows how bubble release is tied to the moving interface rather than prescribed externally. Bubbles exposed along the upper part of the interface enter an unobstructed rising pathway and are rapidly incorporated into the ascending flow, whereas those released lower down are initially hindered by the overlying ice. The global consequence is evident in the evolution of $f$ [Fig.~\ref{Fig-1}(b)]: bubbly cases melt faster over most of the melting process, with stronger acceleration at larger $\alpha_0$. Because the bubbly systems advance more rapidly in time, comparisons at equal time conflate the bubble effect with differences in melt state. We therefore compare $\dot f$ with the bubble-free value $\dot f_0$ at the same $f$.

The state-matched response is strongly nonuniform. Using $Ra_e=g\beta_T\Delta T(fH)^3/(\nu\kappa)=Ra\,f^3$ as a compact measure of the evolving convective state, Fig.~\ref{Fig-1}(c) shows weak enhancement at low $Ra_e$ followed by a rapid activation. At percent-level gas fractions, the state-local enhancement can exceed unity, corresponding to more than a doubling of the melt rate.
This activation shows that an available reservoir of bubbles becomes dynamically effective only once the evolving flow and interface establish a connected pathway through which bubble buoyancy can influence the melt-controlling boundary layer.

\begin{figure*}[!t]
\centering
\includegraphics[width=1\textwidth]{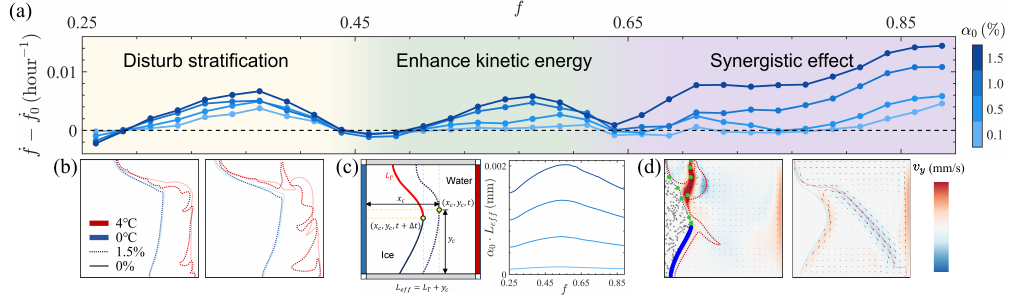}
\caption{\textbf{State dependence of bubble-induced melt acceleration.}
\textbf{(a)} State-matched increment $\Delta\dot f=\dot f-\dot f_0$ versus $f$ for representative gas fractions. The shaded regions identify three stages: disruption of stratification, enhancement of kinetic energy, and their combined action.
\textbf{(b)} Representative melt fronts and $4\,^{\circ}\mathrm{C}$ isotherms in the first stage, comparing bubbly and bubble-free states.
\textbf{(c)} Geometry of the interface turning point $(x_c,y_c)$ and effective release length $L_{\rm eff}=L_\Gamma+y_c$. The right panel shows the evolution of the loading-weighted accessible release measure $\alpha_0L_{\rm eff}$.
\textbf{(d)} Representative late states illustrating the simultaneous action of bubble-driven momentum injection and thermal-structure disruption.}
\label{Fig-2}
\end{figure*}

\textit{Why the forcing is state dependent.---}
The physical origin of this state dependence is resolved in Fig.~\ref{Fig-2}. Across the melt history, the increment $\Delta\dot f=\dot f-\dot f_0$ passes through three reproducible dynamical stages [Fig.~\ref{Fig-2}(a)]. Early on, the bubble-free liquid retains a relatively coherent $4\,^{\circ}\mathrm{C}$ layer associated with the density maximum of freshwater, a structure known to reorganize convective ice melting and its interfacial morphology~\cite{wang2021ice,Weady2022AnomalousConvective}. Released bubbles cross and deform this layer, weaken the stable thermal organization, and move the $4\,^{\circ}\mathrm{C}$ contour toward the melting interface [Fig.~\ref{Fig-2}(b)]. The initial enhancement is therefore controlled primarily by the disruption of thermal stratification.

Once the $4\,^{\circ}\mathrm{C}$ structure is strongly deformed in both bubbly and bubble-free flows, the dominant distinction shifts to the mechanical accessibility of released bubbles. The point $(x_c,y_c)$ farthest from the cold wall separates the unobstructed upper release region from the geometrically hindered lower region. Together with the interface length $L_\Gamma$ above it, this motivates $L_{\rm eff}=L_\Gamma+y_c$ as a measure of accessible release. The evolution of $\alpha_0L_{\rm eff}$ in Fig.~\ref{Fig-2}(c) follows the strengthening of the intermediate-stage response. At later states, the bubble-free system recovers a more coherent thermal structure, whereas rising bubbles sustain stronger upward motion near the melting interface and continued thermal disruption [Fig.~\ref{Fig-2}(d)]. Momentum injection and stratification disruption therefore act together in the late stage. The three stages are different manifestations of the same released-bubble feedback, whose efficiency is controlled jointly by the instantaneous interface geometry and flow state. We therefore turn to the mechanical input emerging from their combined action.

\textit{Mechanical-power criterion.---}
A more transferable description is obtained by comparing the mechanical power supplied by released bubbles with that sustaining the background flow. We therefore define
\begin{equation}
\Pi_B
=
\frac{\text{bubble-generated mechanical power}}
{\text{background-flow mechanical power}} .
\label{eq:PiB}
\end{equation}
This ratio measures the strength of the internally generated bubble forcing relative to the mechanical power already available in the background flow.
For the present simulations, this general power ratio can be evaluated directly from the resolved momentum budget as
\begin{equation}
\Pi_B^{\rm DNS}
=
\frac{\langle P_B^+\rangle}
{\langle P_{T,0}^+\rangle},
\qquad
P_B^+
=
\int_{\Omega_l}
[\bm u\!\cdot\!\bm f_B]_+\,dV ,
\label{eq:PiDNS}
\end{equation}
where $[a]_+\equiv\max(a,0)$ and the brackets denote state-conditioned time integration over equal intervals of $f$. Thus, $\Pi_B^{\rm DNS}$ is the realization of the general power ratio in the present configuration, folding the effects of bubble availability, interface geometry, and flow state into the mechanical input actually delivered to the liquid.

\begin{figure}[t]
\centering
\includegraphics[width=1\columnwidth]{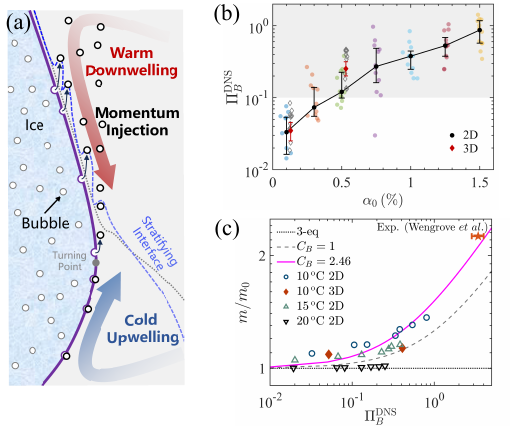}
\caption{\textbf{Mechanical-power criterion and melt-boundary-layer response.}
\textbf{(a)} Schematic of the feedback: bubbles released into a connected rising pathway inject momentum and perturb the stratifying layer, whereas bubbles below the turning point are initially hindered.
\textbf{(b)} State-conditioned power ratio $\Pi_B^{\rm DNS}$ versus $\alpha_0$. Circles and diamonds denote 2D and 3D results, respectively. The gray band $\Pi_B^{\rm DNS}\gtrsim0.1$ marks the approximate mechanically active range in the present simulations.
\textbf{(c)} Relative melt response $m/m_0$ versus $\Pi_B^{\rm DNS}$. The dotted line is the standard three-equation result without bubble forcing, the dashed curve has $C_B=1$, and the magenta curve uses $C_B=2.46$ inferred from the mechanically significant $10\,^{\circ}\mathrm{C}$ 2D cases. The orange star is the independent laboratory estimate from Ref.~\cite{Wengrove2023}.}
\label{Fig-3}
\end{figure}

As $\alpha_0$ rises from $0.1\%$ to $1.5\%$, the median $\Pi_B^{\rm DNS}$ increases from approximately $0.03$ to $0.87$ [Fig.~\ref{Fig-3}(b)]. Bubble effects are weak for $\Pi_B^{\rm DNS}\ll1$ and become clearly non-negligible once the bubble power reaches a substantial fraction of the background power. In the present dataset, this onset occurs around $\Pi_B^{\rm DNS}\sim0.1$. The power ratio quantifies the mechanical forcing supplied by released bubbles, but not how that input maps onto melt enhancement. To express the response on the velocity scale commonly used in melt parameterizations and geophysical observations, we define an equivalent melt velocity $m=H\dot f$, with $m_0=H\dot f_0$ for the bubble-free reference, so that $m/m_0=\dot f/\dot f_0$. We further consider $T_h=15$ and $20\,^{\circ}\mathrm{C}$ to explore progressively stronger background convective states, with all other conditions unchanged.

Figure~\ref{Fig-3}(c) shows that a finite bubble-power input does not imply a unique melt response. The $10\,^{\circ}\mathrm{C}$ and $15\,^{\circ}\mathrm{C}$ cases remain mechanically responsive, with $m/m_0$ increasing with $\Pi_B^{\rm DNS}$, whereas the $20\,^{\circ}\mathrm{C}$ cases stay close to unity despite finite bubble power. This contrast indicates that the effectiveness of bubble-generated mechanical input depends on the state of the receiving flow: sufficiently strong background convection can reduce the fraction of that input that remains dynamically connected to near-interface heat transfer. The resulting bubble-boundary-layer feedback therefore operates over a finite coupling window.
This window reflects two requirements: accessibility of released-bubble forcing to the melt boundary layer and receptivity of that boundary layer to convert the mechanical input into enhanced heat transfer. The activation in Fig.~\ref{Fig-1}(c) and the weakened response in Fig.~\ref{Fig-3}(c) therefore represent two limits of the same coupling process.

\textit{An effective-velocity correction.---}
The distinction between mechanical input and melt response can be made explicit in a form compatible with standard ice-ocean melt closures~\cite{HOLLAND1999ModelingThermodynamic,MCPHEE2008RevisitingHeat,JENKINS2010ObservationParameterization,MCCONNOCHIE2017TestingCommon}. We represent the bubble-generated power as an equivalent increment in the near-boundary mechanical velocity scale,
\begin{equation}
C_DU_{\rm eff}^3=C_DU^3+C_Bg\alpha_{\rm rel}m\ell_B,\qquad
m=M_{3\rm eq}(U_{\rm eff},T,S),
\label{eq:closure}
\end{equation}
where $U$ is the background near-boundary velocity, $U_{\rm eff}$ is its bubble-corrected counterpart, $C_D$ is the drag coefficient, $\alpha_{\rm rel}m$ is the released-gas volume flux per unit ice area, and $\ell_B$ is the effective distance over which bubbles remain mechanically coupled to the boundary layer. The coefficient $C_B$ characterizes the response of the melt boundary layer to bubble-generated mechanical input. For the standard three-equation model with constant exchange coefficients~\cite{HOLLAND1999ModelingThermodynamic,MCPHEE2008RevisitingHeat,JENKINS2010ObservationParameterization,MCCONNOCHIE2017TestingCommon}, $m\propto U$ at fixed far-field $T$, $S$, and pressure, giving
\begin{equation}
\frac{m}{m_0}\simeq\frac{U_{\rm eff}}{U}
=\left(1+C_B\Pi_B\right)^{1/3}.
\label{eq:mratio}
\end{equation}
For the present simulations, $\Pi_B$ is evaluated as $\Pi_B^{\rm DNS}$. The $10\,^{\circ}\mathrm{C}$ data give a median $C_B=2.46$ for mechanically significant cases, decreasing to $1.54$ at $15\,^{\circ}\mathrm{C}$ and $0.17$ at $20\,^{\circ}\mathrm{C}$. Thus $\Pi_B^{\rm DNS}$ measures the realized mechanical input from released bubbles, whereas $C_B$ characterizes how effectively the melt-controlling boundary layer responds to that input. For comparison, the natural ice measurements of Wengrove \textit{et al.}~\cite{Wengrove2023} correspond to $\Pi_B^{\rm lab}\simeq2.86$-$4.0$ and $m/m_0\simeq2.25$, providing an independent laboratory example in which bubble-related mechanical power is comparable to or exceeds the background power.

\begin{figure}[b]
\centering
\includegraphics[width=0.7\columnwidth]{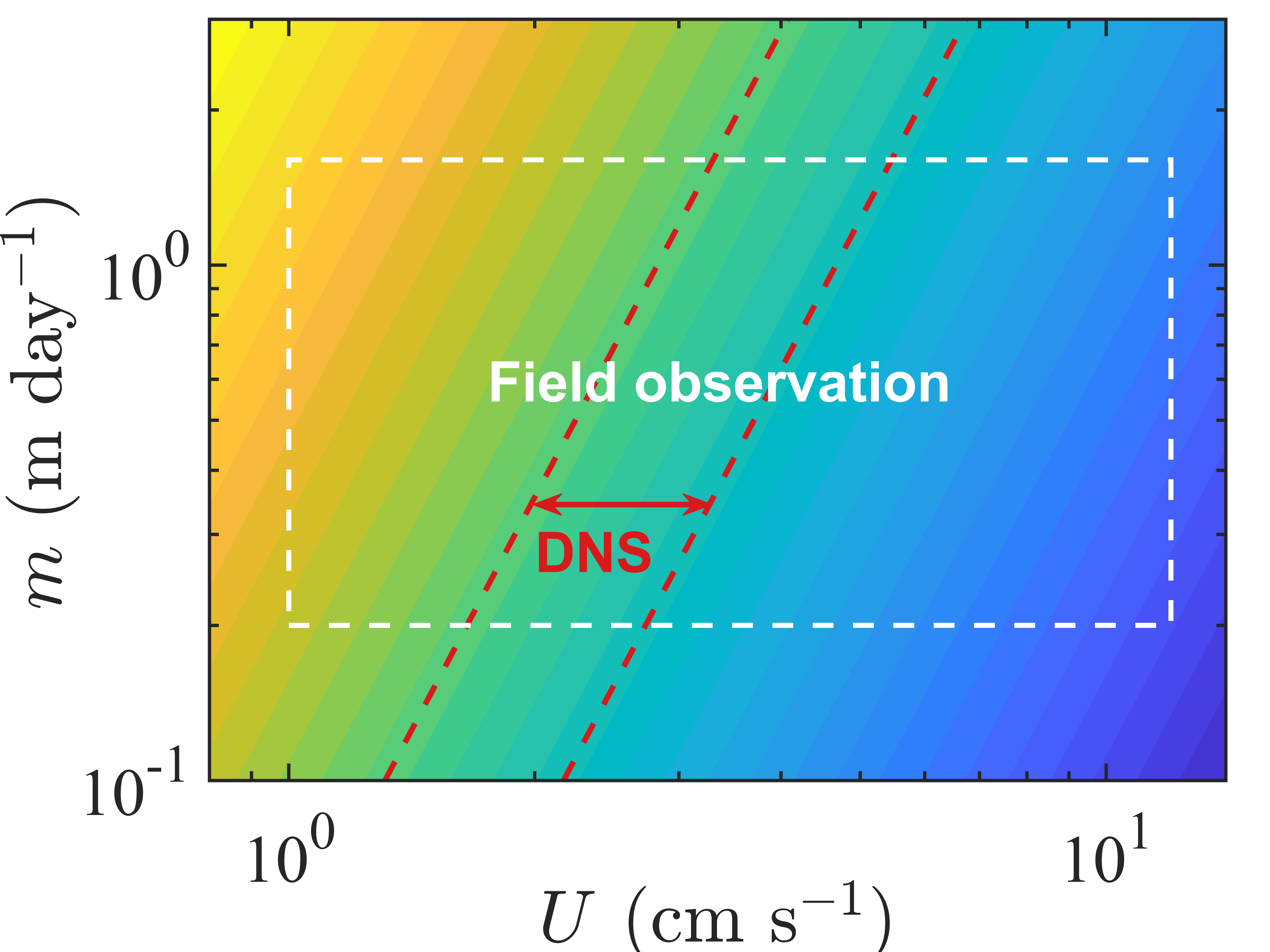}
\caption{\textbf{Scale bridge to observed ice-ocean conditions.}
Power ratio $\Pi_B^{\rm geo}$ in the $(U,m)$ plane using $\alpha_{\rm rel}=1.0\%$, $\ell_B=0.1\,\mathrm{m}$, and $C_D=6\times10^{-3}$. The white dashed rectangle is the reported field-observation range~\cite{WEISS2025DirectMeasurement}. The red dashed contours mark the realized bubble-power range of the $1.0\%$ DNS, $\Pi_B^{\rm DNS}\simeq0.19$-$0.84$.
}
\label{Fig-4}
\end{figure}

\textit{Scale bridge to ice--ocean melting.---}
For an oceanic boundary layer, the corresponding power comparison is between bubble-rise power and shear power. If bubbles released at rate $\alpha_{\rm rel}m$ remain coupled over $\ell_B$, then
\begin{equation}
\Pi_B^{\rm geo}
=\frac{g\alpha_{\rm rel}m\ell_B}{C_DU^3}.
\label{eq:Pigeo}
\end{equation}
Figure~\ref{Fig-4} maps this ratio in the $(U,m)$ plane using $\alpha_{\rm rel}=1.0\%$, $\ell_B=0.1\,\mathrm{m}$, and $C_D=6\times10^{-3}$~\cite{SALGANIK2025ImpactsAir,JENKINS2011ConvectionDrivenMelting,MCCONNOCHIE2017EnhancedAblation}. Within the reported field envelope, $U\simeq1$-$12\,\mathrm{cm\,s^{-1}}$ and $m\simeq0.2$-$1.6\,\mathrm{m\,day^{-1}}$~\cite{WEISS2025DirectMeasurement}, a substantial region yields $\Pi_B^{\rm geo}$ within the mechanically active DNS range $\Pi_B^{\rm DNS}\simeq0.19$-$0.84$ obtained in the $1.0\%$ simulations. This order-of-magnitude comparison shows that, across reported ice-ocean conditions, even modest gas fractions in natural ice can enter a mechanically significant regime relative to the background boundary-layer flow.

A further consequence is that this forcing is not prescribed externally. Because bubble release is tied to the advancing phase boundary, faster melting exposes entrapped gas more rapidly, increasing the available bubble-power flux and feeding back on the near-boundary flow and subsequent melt rate. This self-coupling distinguishes melt-front-released bubbles from externally injected bubbly flows, for which gas flux and injection location can be prescribed independently of phase change~\cite{mathai2020bubbly,gvozdic2018experimental,gvozdic2019experimental}. The feedback remains self-limited by the finite coupling window identified above.

More broadly, the results change the physical role assigned to entrapped air in melting ice. Rather than acting only as a passive material property, entrapped bubbles become a mechanically active dispersed phase when exposed by the moving melt front. The effective-velocity representation in Eq.~(\ref{eq:closure}) provides a minimal route for incorporating this internally activated forcing into existing ice-ocean melt models without altering their freezing-point, heat, or salt balances. The underlying principle extends beyond the present system: a moving phase boundary can activate an embedded dispersed phase as an internal mechanical forcing whenever the released power becomes comparable to that sustaining the boundary-layer flow and remains coupled long enough to modify interfacial transport.

\vspace{20pt}
\noindent
{\bf Acknowledgements:} 
We thank Prof. Roberto Verzicco and Prof. Lian-Ping Wang for insightful suggestions, and Shengqi Zhang, Rui Yang, Yi-Bao Zhang, Lei Ren and Zi-Hao Song for valuable discussions.
This work is supported by the National Natural Science Foundation of China Excellence Research Group Program for ‘Multiscale Problems in Nonlinear Mechanics’ (No.~12588201), the Natural Science Foundation of China under Grant No. 12602337, the National Key R$\&$D Program of China under Grant No. 2025YFF0511800, the Natural Science Foundation of China under Grant No. 12402321, the New Cornerstone Science Foundation through the New Cornerstone Investigator Program and the XPLORER PRIZE, the China Postdoctoral Science Foundation No. 2026M791982, and the Postdoctoral Fellowship Program of CPSF NO. GZC20261056.

\bibliographystyle{apsrev4-2}
\bibliography{Reference}

\appendix

\section{End Matter}
The liquid obeys the incompressible Navier--Stokes equations under the Boussinesq approximation, with freshwater density inversion represented by the same nonlinear equation of state used in the original simulations. Flow and temperature are solved with the \textsc{ch4-project} lattice-Boltzmann framework, while phase change is treated by an enthalpy formulation~\cite{calzavarini2019eulerian,Succi2001LBM,huber2008lattice,2018ESFAHANIBasalMelting,SUGIYAMA2008MicrobubblyDrag}. The reference 2D grid is $512\times512$. The 3D domain uses $512\times512\times32$ nodes with $L_z/H=1/16$ and periodic spanwise boundary conditions. Additional runs at $T_h=15$ and $20\,^{\circ}\mathrm{C}$ vary the background convective state without introducing new bubble physics.

Bubbles are initially embedded in the ice and remain mechanically inactive until the local liquid fraction exposes them to water. For the present diameter $d_B\simeq0.12\,\mathrm{mm}$, $Bo\simeq1.9\times10^{-3}$ and the particle response time is short compared with the resolved convective time. The small-response-time approximation gives~\cite{MAXEY1983EquationMotion,Gatignol1983Faxen}
\begin{equation}
\frac{d\bm x_B}{dt}=\bm u+(\beta-1)\tau_p(D_t\bm u-\bm g),
\qquad \beta\simeq3,
\label{eq:bubble}
\end{equation}
and the corresponding two-way feedback in the light-bubble limit is
\begin{equation}
\bm f_B(\bm x,t)=\rho_0\alpha(\bm x,t)(D_t\bm u-\bm g).
\label{eq:feedback}
\end{equation}
The singular point-particle forcing is regularized over neighboring Eulerian nodes following established point-based two-way-coupling practice~\cite{DEWIT2024EfficientPointbased}. Coupling turns on continuously with the local liquid fraction at the diffuse melt interface. Direct bubble thermal feedback is neglected, because for the present bubble sizes and dilute gas fractions, its contribution to the thermal budget is negligible compared with the mechanical effect of bubble motion~\cite{RAZA2026ParticleThermal}. Bubbles that reach the boundary in the rise direction are removed from subsequent momentum coupling. This removal avoids artificial long-time recirculation in the closed hydrodynamic cavity. Full material properties, regularization, grid tests, and phase-change validation are given in the Supplemental Material.

To compare bubbly and bubble-free systems at equal melt state, the power ratio in Eq.~(\ref{eq:PiDNS}) is evaluated over fixed liquid-fraction intervals. The positive bubble power is
\begin{equation}
P_B^+(t)=\int_{\Omega_l}\left[\bm u(\bm x,t)\cdot\bm f_B(\bm x,t)\right]_+\,dV,
\end{equation}
and the background reference is
\begin{equation}
P_{T,0}^+(t)=\int_{\Omega_l}\left[\bm u_0(\bm x,t)\cdot\bm f_{T,0}(\bm x,t)\right]_+\,dV.
\end{equation}
For $f\in[f_k,f_k+\Delta f]$,
\begin{equation}
\Pi_B^{\rm DNS}(f_k)=
\frac{\int_{t:\,f(t)\in[f_k,f_k+\Delta f]}P_B^+(t)\,dt}
{\int_{t_0:\,f_0(t_0)\in[f_k,f_k+\Delta f]}P_{T,0}^+(t_0)\,dt_0}.
\label{eq:statePower}
\end{equation}
The values in Fig.~\ref{Fig-3} use $\Delta f=0.05$ over $0.3\le f\le0.9$. Hence $\Pi_B^{\rm DNS}$ is a time-integrated, state-conditioned measure of positive mechanical input.

For the effective-velocity bridge, the standard three-equation closure is used with constant turbulent exchange coefficients. At fixed far-field temperature, salinity, and pressure its interfacial solution is homogeneous in the friction velocity, yielding $m\propto U$. We verified this proportionality numerically over the parameter range used for Fig.~\ref{Fig-4}. The derivation, together with the laboratory estimate of $\Pi_B^{\rm lab}$ is given in the Supplemental Material.

\end{document}